\documentclass[preprint,showpacs,preprintnumbers,amsmath,amssymb,superscriptaddress]{revtex4}

\include{ams}
\usepackage{amsmath,amssymb,amsthm}
\usepackage{dcolumn}
\usepackage{bm}
\usepackage{graphicx}

\begin{document}

\title{Electronic properties of double-layer carbon nanotubes}

\author{M. Pudlak}
\email{pudlak@saske.sk}
\affiliation{Institute of Experimental Physics, Slovak Academy of Sciences, Watsonova 47,043 53 Kosice, Slovak Republic}

\author{R. Pincak}
\email{pincak@saske.sk} \affiliation{Institute of Experimental
Physics, Slovak Academy of Sciences, Watsonova 47,043 53 Kosice,
Slovak Republic} \affiliation{Joint Institute for Nuclear Research,
BLTP, 141980 Dubna, Moscow region, Russia}

\pacs{73.63.-b, 73.63.Fg, 73.22.-f}

\date{\today}

\begin{abstract}
The electronic spectra for double-wall zigzag and armchair
nanotubes are found. The influence of nanotube curvatures on the
electronic spectra is also calculated. Our finding that the outer
shell is hole doped by the inner shell is in the difference
between Fermi levels of individual shells which originate from the
different hybridization of $\pi$ orbital. The shift and rotation
of the inner nanotube with respect to the outer nanotube are
investigated. We found stable semimetal characteristics of the
armchair DWNTs in regard of the shift and rotation of the inner
nanotube. We predict the shift of $k_F$ towards the bigger wave
vectors with decreasing of the radius of the armchair nanotube.

\end{abstract}

\maketitle

\section{Introduction}
Carbon nanotubes are very interesting because of their unique
mechanical and electronic properties. A single-wall carbon nanotube
can be described as a graphene sheet rolled into a cylindrical shape
so that the structure is one-dimensional with axial symmetry and in
general exhibiting a spiral conformation called chirality. The
primary symmetry classification of carbon nanotubes is either
achiral (symmorphic) or chiral (non-symmorphic). Achiral carbon
nanotubes are defined by a carbon nanotube whose mirror images have
an identical structure to the original one. There are only two cases
of achiral nanotubes, armchair and zigzag nanotubes. The names of
armchair and zigzag nanotubes arise from the shape of the
cross-section ring at the edge of the nanotubes. Chiral nanotubes
exhibit spiral symmetry whose mirror image cannot be superposed onto
the original one. There is a variety of geometries in carbon
nanotubes which can change the diameter, chirality and cap
structures. The electronic structure of carbon nanotubes is derived
by a simple tight-binding calculation for the $\pi$-electrons of
carbon atoms. Of special interest is the prediction that the
calculated electronic structure of a carbon nanotube can be either
metallic or semiconducting, depending on its diameter and chirality.
The energy gap for a semiconductor nanotube, which is inversely
proportional to its diameters, can be directly observed by scanning
tunneling microscopy measurements. The electronic structure of a
single-wall nanotube can be obtained simply from that of
two-dimensional graphite. By using periodic boundary conditions in
the circumferential direction denoted by the chiral vector $C_{h}$,
the wave vector associated with the $C_{h}$ direction becomes
quantized, while the wave vector associated with the direction of
the translational vector $T$ along the nanotube axis remains
continuous for a nanotube of infinite length. Thus, the energy bands
consist of a set of one-dimensional energy dispersion relations
which are cross sections of those for two-dimensional graphite. To
obtain explicit expressions for the dispersion relations, the
simplest cases to consider are the nanotubes having the highest
symmetry, e.g. highly symmetric achiral nanotubes. The synthesis of
DWNTs has been reported recently~\cite{sugai,zhou}. Their electronic
structure was investigated by the local density
approximation~\cite{kwon,saito,okada,Zolyomi,Kurti} and the
tight-binding model~\cite{ho,dress,lin1,lambin}. A similar method
can be used to investigate the electronic spectra of the fullerene
molecules~\cite{Pudlak1,Pincak2}. In this paper we are interested in
the zigzag and armchair double-wall nanotubes(DWNTs) with a small
radius. In these DWNTs the difference of Fermi levels of individual
nanotubes has to be taken into account. We focus on $(9,0)-(18,0)$
zigzag tubules and $(5,5)-(10,10)$ armchair tubules. They are the
best matched, double layer tubules.

\section{$(9,0)-(18,0)$ zigzag tubules}
Firstly, we describe the model for the zigzag nanotubes. The $\pi $
electronic structures are calculated from the tight-binding
Hamiltonian
$$ H=\sum_{i}\epsilon |\varphi
^{out}_{i}\rangle \langle \varphi ^{out}_{i}| +\sum
_{i,j}\gamma_{ij} \left(|\varphi ^{in}_{i}\rangle \langle \varphi
^{in}_{j}|+h.c\right)+ \sum_{i} \widetilde{\epsilon}|\varphi
^{in}_{i}\rangle \langle \varphi ^{in}_{i}|+\sum
_{i,j}\widetilde{\gamma}_{ij} \left(|\varphi ^{in}_{i}\rangle
\langle \varphi ^{in}_{j}|+h.c\right) $$
\begin{equation}
+\sum_{l,n} W_{ln} \left(|\varphi ^{in}_{l}\rangle \langle \varphi
^{out}_{n}|+h.c\right),
\end{equation}
$\epsilon$ and $\widetilde{\epsilon}$ are Fermi energies of the
outer and inner nanotubes; $|\varphi ^{out}_{i}\rangle$, $|\varphi
^{in}_{i}\rangle$ are $\pi $ orbitals on site $i$ at the outer and
inner tubes; $\gamma_{ij}$, $\widetilde{\gamma}_{ij}$ are the
intratube hopping integrals; $W_{ij}$ are the intertube hoping
integrals which depends on the distance $d_{ij}$ and angle
$\theta_{ij}$ between the $\pi_{i}$ and $\pi_{j}$ orbitals
(see~\cite{roche1,lee, kang} for details).
\begin{equation}
W_{ij}=\frac{\gamma_{0}}{8}\cos(\theta_{ij})e^{(\xi-d_{ij})/\delta},
\end{equation}
where $\theta_{ij}$ is an angle between the \textit{i}th atom of the
inner shell and the \textit{j}th atom of the outer shell, $d_{ij}$
is the interatom distance and $\xi$ is a intertube distance. The
characteristic length $\delta=0.45{\AA}$.

To describe the parameter which characterized the zig-zag tubules,
we start from the graphene layer~\cite{roche} where we can define
the vectors connecting the nearest neighbor carbon atoms for zigzag
nanotubes in the form:
\begin{equation}
\overrightarrow{\tau_{1}}=a(0;\frac{1}{\sqrt{3}}),\nonumber
\end{equation}
\begin{equation}
\overrightarrow{\tau_{2}}=a(\frac{1}{2};-\frac{1}{2\sqrt{3}}),\nonumber
\end{equation}
\begin{equation}
\overrightarrow{\tau_{3}}=a(-\frac{1}{2};-\frac{1}{2\sqrt{3}}).
\end{equation}
\begin{figure}[htb]
\centering
\includegraphics[width=1.4\textwidth]{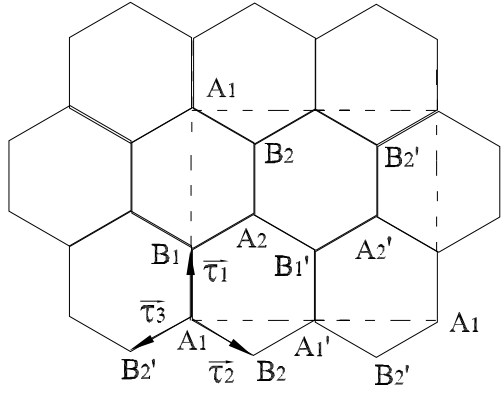}
\caption{The outer shell part of the unit cell in the case of zigzag
nanotubes.}\label{fig3}
\end{figure}
\begin{figure}[htb]
\centering
\includegraphics[width=1.7\textwidth]{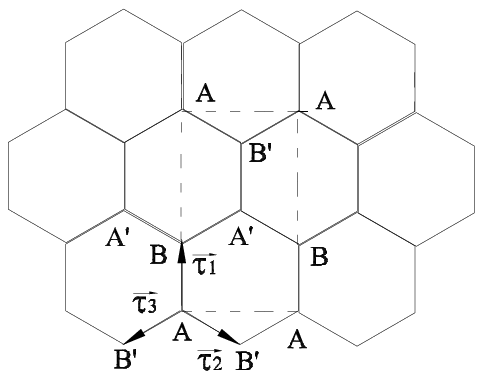}
\caption{The inner shell part of the unit cell in the case of zigzag
nanotubes. }\label{fig4}
\end{figure}
The distance between atoms in the unit cell is
$d=|\overrightarrow{\tau_{i}}|=\frac{a}{\sqrt{3}}$. Following the
scheme in Figs 1,2~\cite{saito1} we want to find solution to the
double-layer graphene tubules in the form:
\begin{equation}
\psi(\overrightarrow{r})=\psi_{out}(\overrightarrow{r})+\psi_{in}(\overrightarrow{r}),
\end{equation}
where
\begin{eqnarray}
\psi_{out}(\overrightarrow{r})=C_{A_{1}}\psi_{A_{1}}+C_{A_{2}}\psi_{A_{2}}
+C_{B_{1}}\psi_{B_{1}}+C_{B_{2}}\psi_{B_{2}}\nonumber\\
+C_{A_{1}^{`}}\psi_{A_{1}^{`}}+C_{A_{2}^{`}}\psi_{A_{2}^{`}}+C_{B_{1}^{`}}\psi_{B_{1}^{`}}+C_{B_{2}^{`}}\psi_{B_{2}^{`}},
\end{eqnarray}
and
\begin{equation}
\psi_{in}(\overrightarrow{r})=C_{A}\psi_{A}+C_{B}\psi_{B}+C_{A^{`}}\psi_{A^{`}}+C_{B^{`}}\psi_{B^{`}}.
\end{equation}
We want to find solution to the above equation in the form of the
Bloch function
\begin{equation}
\psi_{\alpha}(\overrightarrow{k},\overrightarrow{r})=
\frac{1}{\sqrt{M}}\sum_{n}e^{i\overrightarrow{k}(\overrightarrow{r_{n}}
+\overrightarrow{d}_{\alpha})}
|\varphi(\overrightarrow{r}-\overrightarrow{r}_{n}-\overrightarrow{d}_{\alpha})\rangle,
\end{equation}
where $\alpha$ denotes $A$ or $B$ atoms. Here
$\overrightarrow{d}_{\alpha}$ is the coordinate of the $\alpha$ atom
in the unit cell and $\overrightarrow{r_{n}}$ is a position of a
unit cell, $M$ is a number of the unit cell;
$|\varphi(\vec{r})\rangle$ is a $\pi$ orbital which is generally
different for the outer and inner shell. We denote
\begin{equation}
\epsilon=\langle\varphi^{out}(r-A_{i})|H|\varphi^{out}(r-A_{i})\rangle=
\langle\varphi^{out}(r-B_{i})|H|\varphi^{out}(r-B_{i})\rangle,
\end{equation}
\begin{equation}
\widetilde{\epsilon}=\langle\varphi^{in}(r-A_{i})|H|\varphi^{in}(r-A_{i})\rangle=
\langle\varphi^{in}(r-B_{i})|H|\varphi^{in}(r-B_{i})\rangle.
\end{equation}
Now we define the intratube hopping integrals
\begin{equation}
\langle\varphi^{out}(r-A_{1})|H|\varphi^{out}(r-B_{1})\rangle=\gamma_{0},\nonumber
\end{equation}
\begin{equation}
\langle\varphi^{out}(r-A_{1})|H|\varphi^{out}(r-B_{2})\rangle
=\gamma_{0}\beta=\langle\varphi^{out}(r-A_{1})|H|\varphi^{out}(r-B_{2}^{`})\rangle,
\end{equation}
and
\begin{equation}
\langle\varphi^{in}(r-A)|H|\varphi^{in}(r-B)\rangle=\gamma_{0},\nonumber
\end{equation}
\begin{equation}
\langle\varphi^{in}(r-A)|H|\varphi^{in}(r-B^{'})\rangle
=\gamma_{0}\widetilde{\beta},
\end{equation}
where $\gamma_{0}$ is the hoping integral in the graphene and $\beta
(\widetilde{\beta})$ is part which depends on the surface curvature
and will be computed latter. So in a tight-binding approximation we
get the systems of equations as showing in Appendix A.

Firstly, we solve the equations in Appendix A assuming that $W_{ij}$
is the perturbation. So we can decouple these $12$ equations. We get
$8$ equations for the outer shell and $4$ for the inner shell. If we
express the state of the outer shell (Eq.5) in the form
$\psi_{out}=(C_{A_{1}},C_{B_{1}},C_{A_{2}},C_{B_{2}},C_{A_{1}^{`}},C_{B_{1}^{`}},C_{A_{2}^{`}},C_{B_{2}^{`}})$,
we get the solutions to the outer shell in the form
\begin{equation}
E_{1,2}(k)=\epsilon\pm\gamma_{0}(1+4\beta\cos\frac{m\pi}{N}\cos\frac{\sqrt{3}ka}{2}
+4\beta^{2}\cos^{2}\frac{m\pi}{N})^{\frac{1}{2}},\nonumber
\end{equation}
\begin{equation}
\psi_{1,2}=\frac{1}{\sqrt{8}}\left(1;\pm e^{-i\varphi_{1}};1;\pm
e^{-i\varphi_{1}},1;\pm e^{-i\varphi_{1}};1; \pm
e^{-i\varphi_{1}}\right)
\end{equation}
\begin{equation}
E_{3,4}(k)=\epsilon\pm\gamma_{0}(1-4\beta\cos\frac{m\pi}{N}\cos\frac{\sqrt{3}ka}{2}
+4\beta^{2}\cos^{2}\frac{m\pi}{N})^{\frac{1}{2}},\nonumber
\end{equation}
\begin{equation}
\psi_{3,4}=\frac{1}{\sqrt{8}}\left(1;\pm e^{-i\varphi_{2}};-1;\mp
e^{-i\varphi_{2}},1;\pm e^{-i\varphi_{2}};-1;\mp
e^{-i\varphi_{2}}\right)
\end{equation}
\begin{equation}
E_{5,6}(k)=\epsilon\pm\gamma_{0}(1+4\beta\sin\frac{m\pi}{N}\cos\frac{\sqrt{3}ka}{2}+
4\beta^{2}\sin^{2}\frac{m\pi}{N})^{\frac{1}{2}}, \nonumber
\end{equation}
\begin{equation}
\psi_{5,6}=\frac{1}{\sqrt{8}}\left(1;\pm e^{-i\varphi_{3}};-i;\mp
ie^{-i\varphi_{3}},-1;\mp e^{-i\varphi_{3}};i;\pm
ie^{-i\varphi_{3}}\right)
\end{equation}
\begin{equation}
E_{7,8}(k)=\epsilon\pm\gamma_{0}(1-4\beta\sin\frac{m\pi}{N}\cos\frac{\sqrt{3}ka}{2}+
4\beta^{2}\sin^{2}\frac{m\pi}{N})^{\frac{1}{2}}, \nonumber
\end{equation}
\begin{equation}
\psi_{7,8}=\frac{1}{\sqrt{8}}\left(1;\pm e^{-i\varphi_{4}};i;\pm
ie^{-i\varphi_{4}},-1;\mp e^{-i\varphi_{4}};-i;\mp
ie^{-i\varphi_{4}}\right)
\end{equation}
where, for instance,
\begin{equation}
e^{i\varphi_{1}}=\frac{e^{i\frac{ka}{\sqrt{3}}}
+2\beta\cos\frac{m\pi}{N}e^{-i\frac{ka}{2\sqrt{3}}}}{(1+4\beta\cos\frac{m\pi}{N}\cos\frac{\sqrt{3}ka}{2}
+4\beta^{2}\cos^{2}\frac{m\pi}{N})^{\frac{1}{2}}}.
\end{equation}
Similar results for the electronic spectra in the case of inner
nanotubes were found in the form
$(\psi_{in}=(C_{A},C_{B},C_{A^{`}},C_{B^{`}}))$
\begin{equation}
E_{9,10}(k)=\tilde{\epsilon}\pm\gamma_{0}(1+4\widetilde{\beta}\cos\frac{m\pi}{N}\cos\frac{\sqrt{3}ka}{2}
+4\widetilde{\beta}^{2}\cos^{2}\frac{m\pi}{N})^{\frac{1}{2}},
\nonumber
\end{equation}
\begin{equation}
\psi_{9,10}=\frac{1}{\sqrt{4}}\left(1;\pm e^{-i\varphi_{5}};1;\pm
e^{-i\varphi_{5}}\right)
\end{equation}
\begin{equation}
E_{11,12}(k)=\tilde{\epsilon}\pm\gamma_{0}(1-4\widetilde{\beta}\cos\frac{m\pi}{N}\cos\frac{\sqrt{3}ka}{2}
+4\widetilde{\beta}^{2}\cos^{2}\frac{m\pi}{N})^{\frac{1}{2}},
\nonumber
\end{equation}
\begin{equation}
\psi_{11,12}=\frac{1}{\sqrt{4}}\left(1;\pm e^{-i\varphi_{6}};-1;\mp
e^{-i\varphi_{6}}\right).
\end{equation}

Since the radii of the outer and inner nanotubes are different
$\beta\neq\widetilde{\beta}$. Here $k_{y}=k$ and
$-\frac{\pi}{\sqrt{3}a}<k<\frac{\pi}{\sqrt{3}a}$ is the first
Brillouin zone. As we have a curved surface, the local normals on
the neighboring sites are no longer perfectly aligned and this
misorientation also changes the transfer integral. The change can be
calculated using the curvature tensor $b_{\alpha\beta}$
~\cite{frankel}. The result is
\begin{equation}
\frac{\delta
t_{a}}{t}=-\frac{1}{2}b_{\gamma\beta}b_{\alpha}^{\gamma}\tau_{a}^{\beta}\tau_{a}^{\alpha},
\end{equation}
where the only nonzero term is $b_{xx}b_{x}^{x}=1/R^{2}$. So we have
\begin{eqnarray}
\frac{\delta t_{1}}{t}=0,\\
\frac{\delta
t_{2}}{t}=-\frac{1}{2}b_{xx}b_{x}^{x}(\tau_{2}^{x})^{2}=-\frac{1}{2R^{2}}(\tau_{2}^{x})^{2},\\
\frac{\delta
t_{3}}{t}=-\frac{1}{2}b_{xx}b_{x}^{x}(\tau_{3}^{x})^{2}=-\frac{1}{2R^{2}}(\tau_{3}^{x})^{2}.
\end{eqnarray}
With using the unit vectors we have
$(\tau_{2}^{x})^{2}$=$(\tau_{3}^{x})^{2}$=$\frac{a^{2}}{4}$. We
found the radius of the inner nanotube from the expression $2\pi
R=Na$. The nonzero terms are $\frac{\delta t_{2}}{t}=\frac{\delta
t_{3}}{t}=\frac{1}{2}(\frac{\pi}{N})^{2}$. The same holds for the
outer nanotube. The parameters $\beta$, $\widetilde{\beta}$ can be
expressed in the form
\begin{equation}
\widetilde{\beta}=1-\frac{\delta
t_{2}}{t}=1-\frac{1}{2}(\frac{\pi}{9})^{2},
\end{equation}
and
\begin{equation}
\beta=1-\frac{\delta t_{2}}{t}=1-\frac{1}{2}(\frac{\pi}{18})^{2}.
\end{equation}
Now we calculate the values $\epsilon$ and $\tilde{\epsilon}$ which
are different because the inner and outer shell radii are different.
Due to the curvature the coordinates of $\overrightarrow{\tau_{i}}$
in space are
\begin{equation}
\overrightarrow{\tau_{1}}=d(0;1;0),\nonumber
\end{equation}
\begin{equation}
\overrightarrow{\tau_{2}}=d(\frac{\sqrt{3}}{2}\cos\theta;-\frac{1}{2};-\frac{\sqrt{3}}{2}\sin\theta),\nonumber
\end{equation}
\begin{equation}
\overrightarrow{\tau_{3}}=d(-\frac{\sqrt{3}}{2}\cos\theta;-\frac{1}{2};-\frac{\sqrt{3}}{2}\sin\theta),
\end{equation}
where $\sin\theta=a/4R$; $R$ is the radius of the nanotube. Now one
can construct three hybrids along the three directions of the bonds.
These directions are
\begin{equation}
\overrightarrow{e_{1}}=(0;1;0),\nonumber
\end{equation}
\begin{equation}
\overrightarrow{e_{2}}=(\frac{\sqrt{3}}{2}\cos\theta;-\frac{1}{2};-\frac{\sqrt{3}}{2}\sin\theta),\nonumber
\end{equation}
\begin{equation}
\overrightarrow{e_{3}}=(-\frac{\sqrt{3}}{2}\cos\theta;-\frac{1}{2};-\frac{\sqrt{3}}{2}\sin\theta).
\end{equation}
The requirement of the orthonormality of the hybrid wave functions
determines uniquely the fourth hybrid, denoted by $|\pi\rangle $,
which corresponds to the $p_{z}$ orbital in graphite. The
hybridization of the $\sigma$ bonds therefore changes from the
uncurved expression to
\begin{equation}
|\sigma_{1}\rangle=s_{1}|s\rangle+\sqrt{1-s_{1}^{2}}|p_{y}\rangle,\nonumber
\end{equation}
\begin{equation}
|\sigma_{2}\rangle=s_{2}|s\rangle+\sqrt{1-s_{2}^{2}}\left(\frac{\sqrt{3}}{2}\cos\theta|p_{x}\rangle
-\frac{1}{2}|p_{y}\rangle-\frac{\sqrt{3}}{2}\sin\theta|p_{z}\rangle\right),\nonumber
\end{equation}
\begin{equation}
|\sigma_{3}\rangle=s_{3}|s\rangle+\sqrt{1-s_{3}^{2}}\left(-\frac{\sqrt{3}}{2}\cos\theta|p_{x}\rangle
-\frac{1}{2}|p_{y}\rangle-\frac{\sqrt{3}}{2}\sin\theta|p_{z}\rangle\right),\nonumber
\end{equation}
\begin{equation}
|\pi\rangle=D_{1}|s\rangle+D_{2}|p_{x}\rangle
+D_{3}|p_{y}\rangle+D_{4}|p_{z}\rangle.
\end{equation}
The mixing parameters $s_{i},D_{j}$ can be determined by the
orthonormality conditions
$\langle\sigma_{i}|\sigma_{j}\rangle=\delta_{ij}$,
$\langle\pi|\sigma_{i}\rangle=0$,$\langle\pi|\pi\rangle=1$. We get

\begin{equation}
|\sigma_{1}\rangle=\frac{1}{\sqrt{3\cos2\theta}}\
|s\rangle+\sqrt{1-\frac{1}{3\cos2\theta}}\ |p_{y}\rangle,\nonumber
\end{equation}
\begin{equation}
|\sigma_{2}\rangle=\sqrt{\frac{3\cos2\theta-1}{3(\cos2\theta+1)}}\
|s\rangle +\sqrt{\frac{2}{3}}\
\frac{1}{\cos\theta}\left(\frac{\sqrt{3}}{2}\cos\theta|p_{x}\rangle
-\frac{1}{2}|p_{y}\rangle-\frac{\sqrt{3}}{2}\sin\theta|p_{z}\rangle\right),\nonumber
\end{equation}
\begin{equation}
|\sigma_{3}\rangle=\sqrt{\frac{3\cos2\theta-1}{3(\cos2\theta+1)}}\
|s\rangle +\sqrt{\frac{2}{3}}\
\frac{1}{\cos\theta}\left(-\frac{\sqrt{3}}{2}\cos\theta|p_{x}\rangle
-\frac{1}{2}|p_{y}\rangle-\frac{\sqrt{3}}{2}\sin\theta|p_{z}\rangle\right),\nonumber
\end{equation}
\begin{equation}
|\pi\rangle=\tan\theta\
\sqrt{\frac{3\cos2\theta-1}{3\cos2\theta}}|s\rangle+\frac{\tan\theta}{\sqrt{3\cos2\theta}}\
|p_{y}\rangle+\frac{\sqrt{\cos2\theta}}{\cos\theta}\ |p_{z}\rangle.
\end{equation}
Now we can find the expression for the $\pi$ orbital to the lowest
order in $a/R$
\begin{equation}
|\pi\rangle\approx\
\frac{a}{2\sqrt{6}R}|s\rangle+\frac{a}{4\sqrt{3}R}\
|p_{y}\rangle+|p_{z}\rangle,
\end{equation}
and so we get
\begin{equation}
\varepsilon=\langle\pi|H|\pi\rangle\approx\
\frac{a^{2}}{24R^{2}}\langle s|H|s\rangle +
\frac{a^{2}}{48R^{2}}\langle p_{y}|H|p_{y}\rangle+\langle
p_{z}|H|p_{z}\rangle.
\end{equation}
Due to $a/2R=\pi/N$,$(N=9)$ we have
\begin{equation}
\tilde{\epsilon}= \frac{1}{6}\frac{\pi^{2}}{N^{2}}\langle
s|H|s\rangle + \frac{1}{12}\frac{\pi^{2}}{N^{2}}\langle
p_{y}|H|p_{y}\rangle+\langle p_{z}|H|p_{z}\rangle,
\end{equation}
and
\begin{equation}
\epsilon= \frac{1}{24}\frac{\pi^{2}}{N^{2}}\langle s|H|s\rangle +
\frac{1}{48}\frac{\pi^{2}}{N^{2}}\langle
p_{y}|H|p_{y}\rangle+\langle p_{z}|H|p_{z}\rangle.
\end{equation}
In the case $m=3$ we find
\begin{equation}
E_{3,4}(k)=\epsilon\pm\gamma_{0}(1-2\beta\cos\frac{\sqrt{3}ka}{2}
+\beta^{2})^{\frac{1}{2}},
\end{equation}
\begin{equation}
E_{11,12}(k)=\tilde{\epsilon}\pm\gamma_{0}(1-2\widetilde{\beta}\cos\frac{\sqrt{3}ka}{2}
+\widetilde{\beta}^{2})^{\frac{1}{2}},
\end{equation}
where $k=0$ is a Fermi point for both the inner and outer nanotubes
in the case $\beta=\widetilde{\beta}=1$. Nanotubes have no gap and
have a semiconductor character. If we impose a curvature correction,
we get a gap
\begin{equation}
E_{g}=2(1-\beta)=\gamma_{0}
\left(\frac{\pi}{2N}\right)^{2}=\frac{\gamma_{0}}{4}\left(
\frac{a}{2R}\right)^{2},
\end{equation}
for the outer nanotube and
\begin{equation}
E_{g}=2(1-\widetilde\beta)=\gamma_{0}
\left(\frac{\pi}{N}\right)^{2}=\frac{\gamma_{0}}{4}\left(
\frac{a}{R}\right)^{2},
\end{equation}
for the inner nanotube. Here $R$ is the radius of the inner tube and
$2R$ is the radius of the outer tube. So we get the same gap as was
computed in~\cite{kleiner} where the rehybridized orbital method was
used. For $\gamma_{0}\approx 3\ eV $ we get $E_{g}\approx 0.365 \
eV$ for the inner tube and $E_{g}\approx 0.091 \ eV$ for the outer
tube. Now we want to estimate the difference between "Fermi levels"
of the inner and the outer shell. We have ~\cite{lomer}
\begin{equation}
\langle s|H|s\rangle \approx -12eV,
\end{equation}
\begin{equation}
\langle p_{y}|H|p_{y}\rangle \approx -4eV,
\end{equation}
and the difference is
\begin{equation}
\epsilon-\tilde{\epsilon}=\frac{1}{6}\left(\left(\frac{\pi}{2N}\right)^{2}
-\left(\frac{\pi}{N}\right)^{2}\right)\langle s|H|s\rangle +
\frac{1}{12}\left(\left(
\frac{\pi}{2N}\right)^{2}-\left(\frac{\pi}{N}\right)^{2}\right)\langle
p_{y}|H|p_{y}\rangle.
\end{equation}
From the expression above we finally get the value for the energy
gap
\begin{equation}
\epsilon-\tilde{\epsilon}\approx 0.21eV.
\end{equation}
Now we use the eigenstates $\psi_{i}$ to find the solution when the
interaction between shells is imposed. We assume the symmetric
geometry of zig-zag DWNT. It means that the atoms $A$,$A_{1}$ and
$B$,$B_{1}$ are directly one above another in the neighboring
shells~\cite{lin1}. We take into account only the interactions
\begin{equation}
W_{A,A_{1}}=W_{B,B_{1}}=\frac{\gamma_{0}}{8}.
\end{equation}
We look for solution in the form
\begin{equation}
\Psi=\sum_{i=1}^{12} \zeta_{i}\psi_{i}.
\end{equation}
We have secular equations
\begin{equation}
\sum_{j=1}^{12} \langle \psi_{i}|H|\psi_{j}\rangle
\zeta_{j}=\tilde{E} \zeta_{i},
\end{equation}
where
\begin{equation}
\langle \psi_{i}|H|\psi_{j}\rangle =\delta _{ij}E_{i},
\end{equation}
for $i,j=1,...8$ and $i,j=9,...12$, and the interaction between
shells is described by the terms $ \langle
\psi_{i}|H|\psi_{j}\rangle$ for $i=1,...8$ ; $j=9,...12$ and vice
versa. We have, for instance,
\begin{equation}
\langle
\psi_{9}|H|\psi_{1}\rangle=\frac{1}{4\sqrt{2}}\frac{\gamma_{0}}{8}\left(1+e^{i(\varphi_{5}-\varphi_{1})}\right),
\end{equation}
\begin{equation}
\langle
\psi_{9}|H|\psi_{2}\rangle=\frac{1}{4\sqrt{2}}\frac{\gamma_{0}}{8}\left(1-e^{i(\varphi_{5}-\varphi_{1})}\right).
\end{equation}
We get the eigenvalues $\tilde{E}_{i}$ with eigenvectors which can
be expressed in the form
\begin{equation}
\Psi_{i}=\sum_{j=1}^{12} \zeta_{i,j}\psi_{j}.
\end{equation}
The eigenvalues of Eq.(55) for some values of $\sqrt{3}ka/2$ near
the point $k=0$  are depicted on Fig.3 where $E_c$ and $E_v$ are
conductive and valence band. The band structure for zig-zag DWNT
without intertube interactions is also shown for comparison (Fig.4).
\begin{figure}[htb]
  \centering
\includegraphics[width=0.44\textwidth]{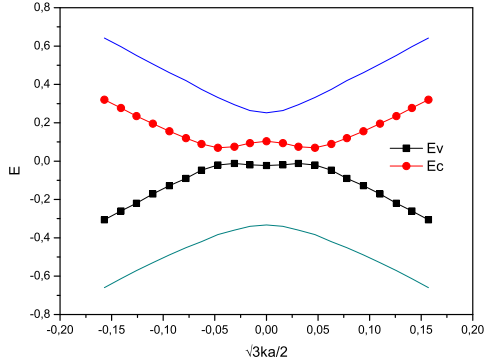}
\caption{Spectra of zigzag DWNT with the intertube interactions.}
\end{figure}
\begin{figure}[htb]
  \centering
\includegraphics[width=0.44\textwidth]{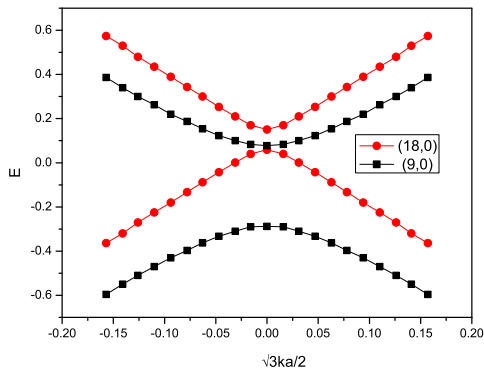}
\caption{Spectra of zigzag DWNT in the absence of the intertube
interactions.}
\end{figure}
At the point k=0 we get the wave function of the valence band
\begin{equation}
\Psi_{v}\simeq -0.6\psi_{3} + 0.8\psi_{11}.
\end{equation}
$\psi_{3}$($\psi_{11}$) is $\pi^{*}$ state of the outer(inner)
nanotube. We get a minimum gap $E_{g}\simeq 90$ meV between the
valence and conductive band of the DWNTs at the wave vectors
$\sqrt{3}ka/2\simeq \pm 0.05$. At these points the wave function has
the form
\begin{equation}
\Psi_{v}\simeq -0.263i\psi_{3}
+0.838\psi_{4}-(0.14+0.45i)\psi_{11}+(0.29-0.09i)\psi_{12}.
\end{equation}
$\psi_{4}$($\psi_{12}$) is $\pi$ state of the outer(inner)
nanotube. We calculated also the electronic structure of
$(8,0)-(16,0)$ and $(10,0)-(20,0)$ DWNTs. The energy gaps are
collected in the Tables I and II. To compute the gaps $\Delta$ we
used formula (19). The gaps denoted by $\Delta_{KM}$ are computed
with formula used in~\cite{kane}; $\Delta_{TB}$ are gaps
calculated in the simple zone folding tight-binding approximation
where the curvature effects are not taken into account. We compare
our results with the previous computed energy gaps. For
$(8,0)-(16,0)$ DWNTs we get a gap which is significantly greater
than that computed by density functional theory (DFT). It is
mainly caused by that the tight-bounding method gives greater gaps
for nanotubes with a very small diameter than the DFT
computations. Another reason is that we describe DWNTs as one
unified system where single nanotubes partially lose their
individual characteristics due to the interactions. For
$(10,0)-(20,0)$ DWNTs we get a similar gap as in~\cite{dyachkov}.

\begin{table}[htb]
\begin{center}
\begin{tabular}{l c c c c c c c }\hline\hline
\bfseries\bfseries \quad \quad & $SWNT$ & \bfseries $\Delta$ &
\bfseries $\Delta_{KM}$
& \bfseries $\Delta_{TB}$& \bfseries $\Delta_{DFT}$\\
\hline
&$(8,0)$&1.752&1.496&1.42&0.59\\
&$(9,0)$&0.37&0.093&0&0.096\\
&$(10,0)$&0.705&0.966&1.07&0.77\\
&$(16,0)$&0.538&0.634&0.67&0.54\\
&$(18,0)$&0.091&0.023&0&0.013\\
&$(20,0)$&0.619&0.568&0.56&0.50\\
\hline\hline
\end{tabular}
\end{center}
\caption{{\footnotesize The values of the minimum energy gaps for
different types of zig-zag SWNTs. The values are calculated in
$eV$. $\Delta$ and $\Delta_{KM}$ are the gaps computed in the
present paper. The values for comparison $\Delta_{TB}$,
$\Delta_{DFT}$ are computations from simple zone tight-binding and
density functional theory~\cite{Zolyomi1}. }} \label{tab}
\begin{center}
\begin{tabular}{l c c c c c }\hline\hline
\bfseries\bfseries \quad \quad & $DWNT$ & \bfseries $\Delta$
& \bfseries $\Delta_{KM}$& \bfseries $\Delta_{DFT}$\\
\hline
&$(8,0)-(16,0)$&1.234&1.080&0.35\\
&$(9,0)-(18,0)$&0.09&0.061&-\\
&$(10,0)-(20,0)$&0.494&0.676&-\\
\hline\hline
\end{tabular}
\end{center}
\caption{{\footnotesize The values of the minimum energy gaps for
different types of zig-zag DWNTs. The values are calculated in
$eV$, $\Delta_{DFT}$ is taken from ~\cite{song} }} \label{tab}
\end{table}

\newpage

\section{$(5,5)-(10,10)$ armchair tubules}
We can make similar calculations of electronic spectra also in the
case of armchair double-layer nanotubes. The system is characterized
by the same Hamiltonian as in the previous section. We can define
the vectors connecting the nearest neighbor carbon atoms for
armchair nanotubes in the form:
\begin{equation}
\overrightarrow{\tau_{1}}=a(\frac{1}{\sqrt{3}}; 0),\nonumber
\end{equation}
\begin{equation}
\overrightarrow{\tau_{2}}=a(-\frac{1}{2\sqrt{3}};
-\frac{1}{2}),\nonumber
\end{equation}
\begin{equation}
\overrightarrow{\tau_{3}}=a(-\frac{1}{2\sqrt{3}}; \frac{1}{2}).
\end{equation}
The distance between atoms in the unit cell is also
$|\overrightarrow{\tau_{i}}|=\frac{a}{\sqrt{3}}$.

\begin{figure}[htb]
\centering
\includegraphics[width=0.5\textwidth]{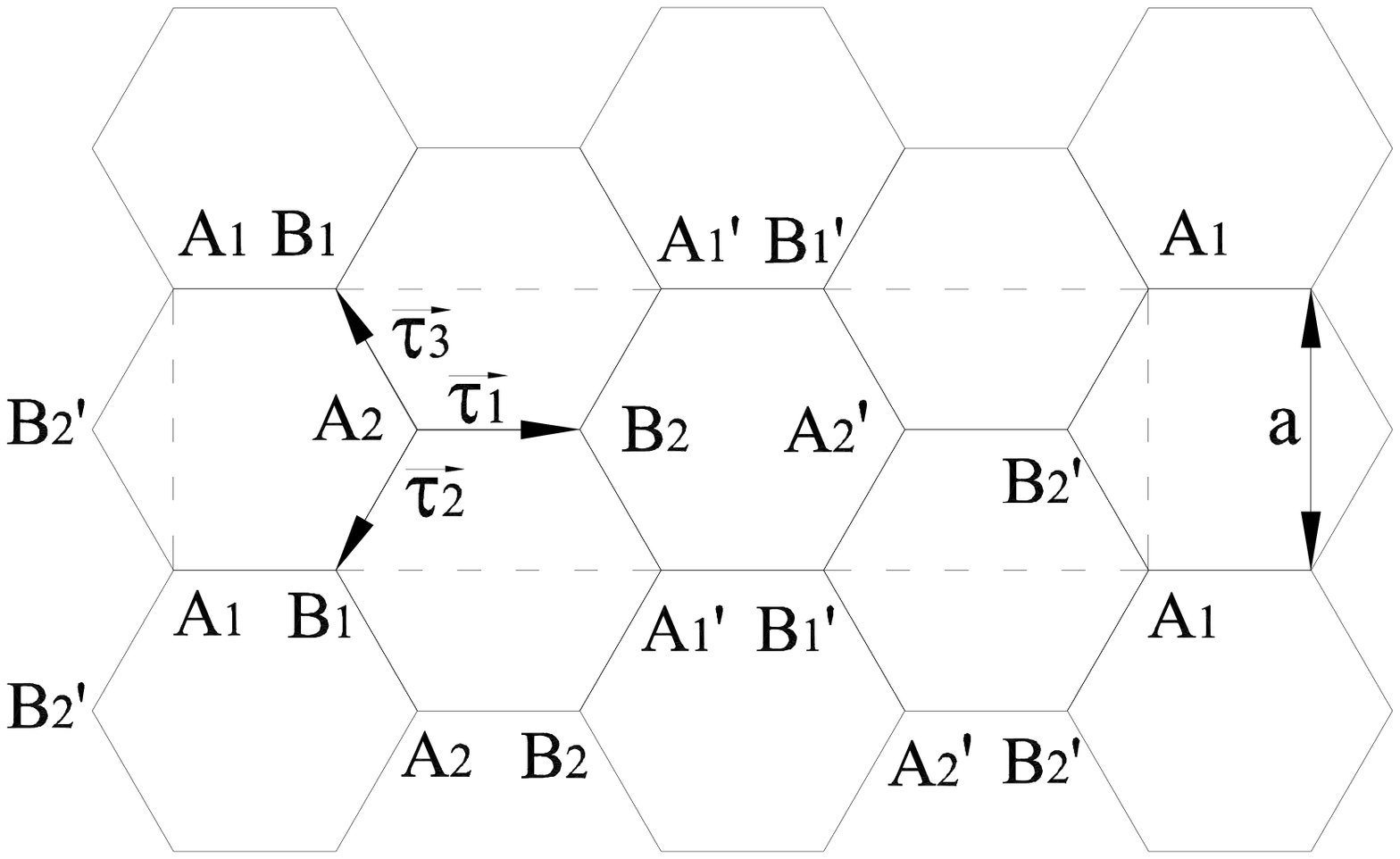}
\caption{The outer shell part of the unit cell in the case of
armchair nanotubes.}\label{fig1}
\end{figure}
\begin{figure}[htb]
\centering
\includegraphics[width=0.9\textwidth]{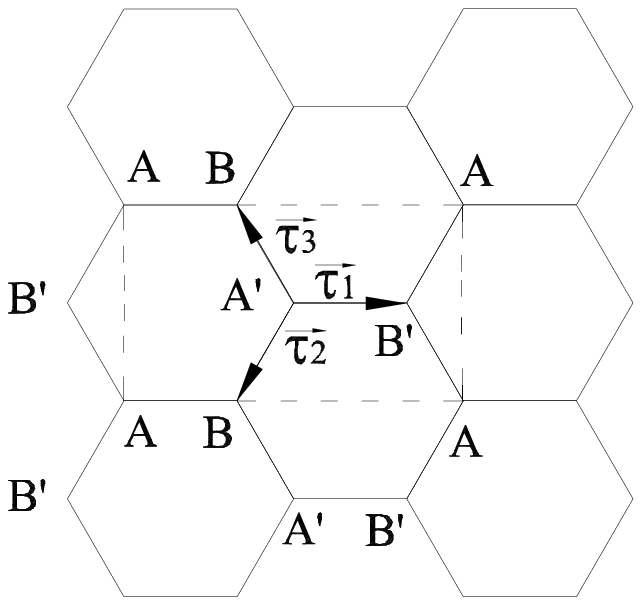}
\caption{The inner shell part of the unit cell in the case of
armchair nanotubes. }\label{fig2}
\end{figure}

Now we define the intratube hopping integrals
\begin{equation}
\langle\varphi^{out}(r-A_{1})|H|\varphi^{out}(r-B_{1})\rangle=\gamma_{0}\alpha,\nonumber
\end{equation}
\begin{equation}
\langle\varphi^{out}(r-A_{1})|H|\varphi^{out}(r-B_{2}^{`})\rangle
=\gamma_{0}\beta,
\end{equation}
and
\begin{equation}
\langle\varphi^{in}(r-A)|H|\varphi^{in}(r-B)\rangle=\gamma_{0}\widetilde{\alpha},\nonumber
\end{equation}
\begin{equation}
\langle\varphi^{in}(r-A)|H|\varphi^{in}(r-B^{'})\rangle
=\gamma_{0}\widetilde{\beta},
\end{equation}
where $\gamma_{0}$ is the hoping integral in the graphene and $
\alpha(\widetilde{\alpha})$, $\beta (\widetilde{\beta})$ are
parameters which describe the dependence of hopping integrals on the
surface curvature. From Figures 5 and 6 we get the system of
equations as describing in Appendix B.

At the beginning we neglect the intertube interactions in the
equations described in Appendix B. We get a set of equations which
can be decoupled. One set for the outer shell and the other for the
inner shell. The electronic spectra and eigenstate for the outer
shell can be expressed in the form
\begin{equation}
E_{1,2}(k)=\epsilon\pm\gamma_{0}(\alpha^{2}
+4\alpha\beta\cos\frac{m\pi}{5}\cos\frac{ka}{2}+4\beta^{2}\cos^{2}\frac{ka}{2})^{\frac{1}{2}},\nonumber
\end{equation}
\begin{equation}
\psi_{1,2}=\frac{1}{\sqrt{8}}\left(1;\pm e^{-i\varphi_{1}};1;\pm
e^{-i\varphi_{1}},1;\pm e^{-i\varphi_{1}};1;\pm
e^{-i\varphi_{1}}\right)
\end{equation}
\begin{equation}
E_{3,4}(k)=\epsilon\pm\gamma_{0}(\alpha^{2}-4\alpha\beta\cos\frac{m\pi}{5}\cos\frac{ka}{2}
+4\beta^{2}\cos^{2}\frac{ka}{2})^{\frac{1}{2}},\nonumber
\end{equation}
\begin{equation}
\psi_{3,4}=\frac{1}{\sqrt{8}}\left(1;\pm e^{-i\varphi_{2}};-1;\mp
e^{-i\varphi_{2}},1;\pm e^{-i\varphi_{2}};-1;\mp
e^{-i\varphi_{2}}\right)
\end{equation}
\begin{equation}
E_{5,6}(k)=\epsilon\pm\gamma_{0}(\alpha^{2}+4\alpha\beta\sin\frac{m\pi}{5}\cos\frac{ka}{2}
+4\beta^{2}\cos^{2}\frac{ka}{2})^{\frac{1}{2}},\nonumber
\end{equation}
\begin{equation}
\psi_{5,6}=\frac{1}{\sqrt{8}}\left(1;\pm e^{-i\varphi_{3}};-i;\mp
ie^{-i\varphi_{3}},-1;\mp e^{-i\varphi_{3}};i;\pm
ie^{-i\varphi_{3}}\right)
\end{equation}
\begin{equation}
E_{7,8}(k)=\epsilon\pm\gamma_{0}(\alpha^{2}-4\alpha\beta\sin\frac{m\pi}{5}\cos\frac{ka}{2}
+4\beta^{2}\cos^{2}\frac{ka}{2})^{\frac{1}{2}},\nonumber
\end{equation}
\begin{equation}
\psi_{7,8}=\frac{1}{\sqrt{8}}\left(1;\pm e^{-i\varphi_{4}};i;\pm
ie^{-i\varphi_{4}},-1;\mp e^{-i\varphi_{4}};-i;\mp
ie^{-i\varphi_{4}}\right)
\end{equation}
The electronic spectra for the inner nanotubes was found in the form
\begin{equation}
E_{9,10}(k)=\tilde{\epsilon}\pm\gamma_{0}(\widetilde{\alpha}^{2}
+4\widetilde{\alpha}\widetilde{\beta}\cos\frac{m\pi}{5}\cos\frac{ka}{2}
+4\widetilde{\beta}^{2}\cos^{2}\frac{ka}{2})^{\frac{1}{2}},\nonumber
\end{equation}
\begin{equation}
\psi_{9,10}=\frac{1}{\sqrt{4}}\left(1;\pm e^{-i\varphi_{5}};1;\pm
e^{-i\varphi_{5}}\right)
\end{equation}
\begin{equation}
E_{11,12}(k)=\tilde{\epsilon}\pm\gamma_{0}(\widetilde{\alpha}^{2}-4\widetilde{\alpha}\widetilde{\beta}\cos\frac{m\pi}{5}\cos\frac{ka}{2}
+4\widetilde{\beta}^{2}\cos^{2}\frac{ka}{2})^{\frac{1}{2}},\nonumber
\end{equation}
\begin{equation}
\psi_{11,12}=\frac{1}{\sqrt{4}}\left(1;\pm e^{-i\varphi_{6}};-1;\mp
e^{-i\varphi_{6}}\right).
\end{equation}

From the boundary condition $k_{x}L=2\pi m$, $L=N3d$ where
$d=a/\sqrt{3}$ is the nearest neighbor bond length we get
$k_{x}=\frac{2\pi m}{3dN}=\frac{2\pi m}{\sqrt{3}Na}$,
$m=0,1,...N-1$; $3d$ is the length of the unit cell in the
x-direction. Here $k_{y}=k$ and $-\frac{\pi}{a}<k<\frac{\pi}{a}$ is
the first Brillouin zone. In this case, we assume that $N=5$ for the
above spectrum. The value for the parameter $\widetilde{\alpha}$ and
$\widetilde{\beta}$ can be found from the expressions
$\widetilde{\alpha}$=$1-\frac{1}{2}b_{xx}b_{x}^{x}(\tau_{1}^{x})^{2}$=1-$\frac{1}{2R^{2}}\frac{a^{2}}{3}$
and
$\widetilde{\beta}$=$1-\frac{1}{2}b_{xx}b_{x}^{x}(\tau_{2}^{x})^{2}$=1-$\frac{1}{2R^{2}}\frac{a^{2}}{12}
$. The radius for the inner, outer nanotube can be found from the
expressions $2\pi R=N3d=\sqrt{3}Na$, $2\pi R=N6d$, respectively. Now
we make a correction of transfer integral caused by the curvature of
nanotubes
\begin{equation}
\widetilde{\beta}=1-\frac{1}{2}(\frac{\pi}{3N})^{2};\quad
\beta=1-\frac{1}{8}(\frac{\pi}{3N})^{2},
\end{equation}
\begin{equation}
\widetilde{\alpha}=1-2(\frac{\pi}{3N})^{2};\quad
\alpha=1-\frac{1}{2}(\frac{\pi}{3N})^{2}.
\end{equation}
We calculate the values $\epsilon$ and $\tilde{\epsilon}$. Due to
the curvature the coordinates of $\overrightarrow{\tau_{i}}$ in
space are
\begin{equation}
\overrightarrow{\tau_{1}}=d(\cos\theta;0;-\sin\theta),\nonumber
\end{equation}
\begin{equation}
\overrightarrow{\tau_{2}}=d(-\frac{1}{2}\cos\vartheta;-\frac{\sqrt{3}}{2};-\frac{1}{2}\sin\vartheta),\nonumber
\end{equation}
\begin{equation}
\overrightarrow{\tau_{3}}=d(-\frac{1}{2}\cos\vartheta;
\frac{\sqrt{3}}{2}; -\frac{1}{2}\sin\vartheta),
\end{equation}
where $\sin\theta=d/2R$ and $\sin\vartheta=d/4R$; $R$ is the radius
of the nanotube. In a similar way, as in the previous section, we
get

\begin{equation}
|\sigma_{1}\rangle=\frac{\cos(\theta+\vartheta)}{\sqrt{2+\cos^{2}(\theta+\vartheta)}}\
|s\rangle+\sqrt{\frac{2}{2+\cos^{2}(\theta+\vartheta)}}\
\left(\cos\theta|p_{x}\rangle-\sin\theta|p_{z}\rangle\right),\nonumber
\end{equation}
\begin{equation}
|\sigma_{2}\rangle=\frac{1}{\sqrt{3}}\ |s\rangle
+\sqrt{\frac{2}{3}}\ \left(-\frac{1}{2}\cos\vartheta|p_{x}\rangle
-\frac{\sqrt{3}}{2}|p_{y}\rangle-\frac{1}{2}\sin\vartheta|p_{z}\rangle\right),\nonumber
\end{equation}
\begin{equation}
|\sigma_{3}\rangle=\frac{1}{\sqrt{3}}\ |s\rangle
+\sqrt{\frac{2}{3}}\ \left(-\frac{1}{2}\cos\vartheta|p_{x}\rangle
+\frac{\sqrt{3}}{2}|p_{y}\rangle-\frac{1}{2}\sin\vartheta|p_{z}\rangle\right),\nonumber
\end{equation}
$$
|\pi\rangle=\sqrt{\frac{2}{3}}\frac{\sin(\theta+\vartheta)}{\sqrt{2+\cos^{2}(\theta+\vartheta)}}|s\rangle
+\frac{2\sin\theta-\sin\vartheta
\cos(\theta+\vartheta)}{\sqrt{6+3\cos^{2}(\theta+\vartheta)}}|p_{x}\rangle+$$
\begin{equation}
\frac{2\cos\theta+\cos\vartheta
\cos(\theta+\vartheta)}{\sqrt{6+3\cos^{2}(\theta+\vartheta)}}|p_{z}\rangle.
\end{equation}
Now we can find the expression for the $\pi$ orbital to the lowest
order in $d/R$
\begin{equation}
|\pi\rangle\approx\
\frac{\sqrt{2}d}{4R}|s\rangle+\frac{d}{4R}\
|p_{x}\rangle+|p_{z}\rangle.
\end{equation}
Due to $3d N=2\pi R$ we get
\begin{equation}
|\pi\rangle\approx\ \frac{\sqrt{2}\pi}{6N}|s\rangle+\frac{\pi}{6N}\
|p_{x}\rangle+|p_{z}\rangle,
\end{equation}
and so
\begin{equation}
\varepsilon=\langle\pi|H|\pi\rangle\approx\
\frac{1}{18}\left(\frac{\pi}{N}\right)^{2}\langle s|H|s\rangle +
\frac{1}{36}\left( \frac{\pi}{N}\right)^{2}\langle
p_{x}|H|p_{x}\rangle+\langle p_{z}|H|p_{z}\rangle.
\end{equation}
From this expression we derive, if $(N=5)$,
\begin{equation}
\tilde{\epsilon}=\frac{1}{18}\left(\frac{\pi}{N}\right)^{2}\langle
s|H|s\rangle + \frac{1}{36}\left( \frac{\pi}{N}\right)^{2}\langle
p_{x}|H|p_{x}\rangle+\langle p_{z}|H|p_{z}\rangle,
\end{equation}
and
\begin{equation}
\epsilon=\frac{1}{18}\left(\frac{\pi}{2N}\right)^{2}\langle
s|H|s\rangle + \frac{1}{36}\left( \frac{\pi}{2N}\right)^{2}\langle
p_{x}|H|p_{x}\rangle+\langle p_{z}|H|p_{z}\rangle.
\end{equation}
The energy levels $E_{3,4}$ and $E_{11,12}$ define the Fermi point
for $m=0$. We have
\begin{equation}
E_{3,4}(k)=\epsilon\pm\gamma_{0} |\alpha-2\beta\cos\frac{ka}{2}|,
\end{equation}
\begin{equation}
E_{11,12}(k)=\tilde{\epsilon}\pm\gamma_{0}|\widetilde{\alpha}-2\widetilde{\beta}\cos\frac{ka}{2}|,
\end{equation}
and the Fermi point is defined by the equations
\begin{equation}
\widetilde{\alpha}-2\widetilde{\beta}\cos\frac{ka}{2}=0,
\end{equation}
for the inner shell, and
\begin{equation}
\alpha-2\beta\cos\frac{ka}{2}=0,
\end{equation}
for the outer shell, respectively. By virtue of $\beta \geq
\alpha(\tilde{\beta} \geq \tilde{\alpha})$ the curvature does not
open a gap in the case of single nanotubes. Using the values
$\langle s|H|s\rangle \approx -12eV$and $ \langle
p_{x}|H|p_{x}\rangle \approx -4eV$ in the following expression:
\begin{equation}
\epsilon-\tilde{\epsilon}=\frac{1}{18}\left(\left(\frac{\pi}{2N}\right)^{2}
-\left(\frac{\pi}{N}\right)^{2}\right)\langle s|H|s\rangle +
\frac{1}{36}\left(\left(
\frac{\pi}{2N}\right)^{2}-\left(\frac{\pi}{N}\right)^{2}\right)\langle
p_{x}|H|p_{x}\rangle,
\end{equation}
we find
\begin{equation}
\epsilon-\tilde{\epsilon}\approx 0.23 \ eV.
\end{equation}
Now we use the eigenstates $\psi_{i}$ to find the solution when the
interaction between shells is imposed. Similarly, as in the previous
case, we look for the solution in the form
\begin{equation}
\Psi=\sum_{i=1}^{12} \zeta_{i}\psi_{i}.
\end{equation}
We have secular equations
\begin{equation}
\sum_{j=1}^{12} \langle \psi_{i}|H|\psi_{j}\rangle
\zeta_{j}=\tilde{E}\zeta_{i},
\end{equation}
We take into account all intertube interactions between atoms which
have a distance $d_{ij}$ less than $4.2 {\AA}$ similarly as
in~\cite{lee,kang}. We use the value $\xi=3.466$ for the intertube
distance in the numerical computations. We compute spectra for three
different geometries. The first case was symmetric geometry where
the atoms $B_{2}^{`}$($A_{2}$) occupy a position directly above
$A^{`}$($B^{`}$),respectively. In the second case, we shift the
inner shell axially by $0.612 {\AA}$ and in the third case, we
rotate the inner shell by $6^{o}$ from the symmetric orientation. We
get the eigenvalues $\tilde{E}_{i}$ with eigenvectors which can be
expressed in the form
\begin{equation}
\Psi_{i}=\sum_{j=1}^{12} \zeta_{i,j}\psi_{j}.
\end{equation}
The spectra for some values of $ka/2$ near the Fermi points of
single nanotubes are depicted on Figure 7. The point $ka/2=1.086$ is
the Fermi point of the isolated inner nanotube. The point
$ka/2=1.057$ is the Fermi point of the isolated outer nanotube.
Approximately, from point $ka/2=1.054$ to point $1.095$ the
$\tilde{E}_{11}$ levels are below the $\tilde{E}_{4}$ level. So in
the armchair DWNT the state $\Psi_{11}$ is occupied at these points.
The state $\Psi_{11}$ is some mixture of the states $\psi_{i}$. For
example, for the point $ka/2=1.083$ we have that the main part of $
\Psi_{11}$ is $\psi_{11}$ which is $\pi^{*}$ state of the inner
tube.

We get that electrons which are localized in the outer nanotubes
in the case without interaction between shells (or in the case of
single nanotubes) are now localized in the inner nanotubes in the
state which is unoccupied in the single nanotubes. Figures, 8-10
describe how the shift and rotation of the inner nanotube,
similarly as in~\cite{lee}, influence the energy gap between
conductance and valence bands in the DWNTs armchair nanotube where
$E_c$ and $E_v$ are conductive and valence bands. We get similar
results for $(4,4)-(8,8)$ and $(6,6)-(12,12)$ DWNTs.

\begin{figure}[htb]
  \centering
\includegraphics[width=0.44\textwidth]{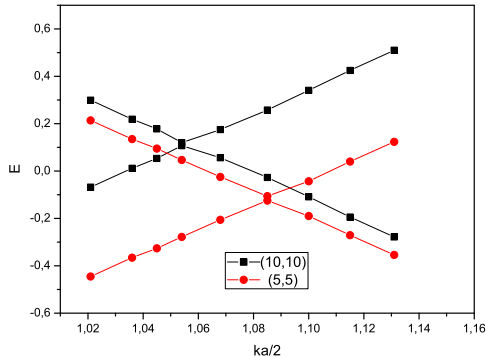}
\caption{Spectra of armchair DWNT in the absence of the intertube
interactions.}
\end{figure}
\begin{figure}[htb]
  \centering
\includegraphics[width=0.44\textwidth]{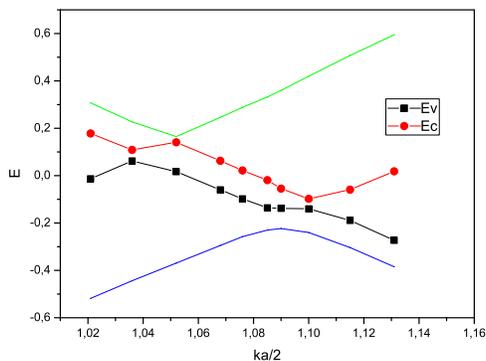}
\caption{Spectra of armchair DWNT with the intertube interactions in
symmetric case.}
\end{figure}
\begin{figure}[htb]
  \centering
\includegraphics[width=0.44\textwidth]{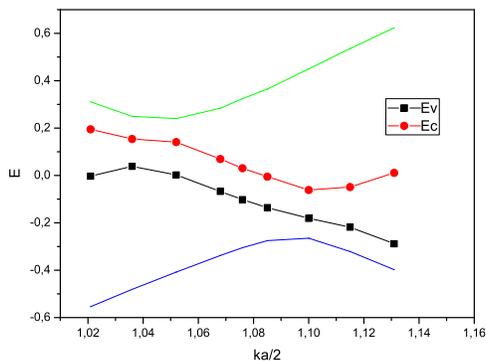}
\caption{Spectra of armchair DWNT with the intertube interactions
with shift of y-axes of inner tube about $\sqrt{3}b/4$ ${\AA}$.}
\end{figure}
\begin{figure}[htb]
  \centering
\includegraphics[width=0.44\textwidth]{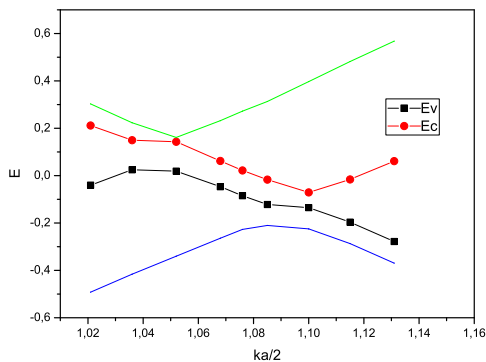}
\caption{ Spectra of armchair DWNT with the intertube interactions
with rotation of inner tube about 6 angle.}
\end{figure}

\newpage

\section{Conclusion}
In the present work, we take into account that the Fermi levels of
the individual nanotubes which create the double wall nanotubes are
different. This difference is very important in the double wall
nanotubes with small diameters. The interplay between energy
difference of the Fermi levels of the individual nanotubes and the
energy gap between valence and conducting band of individual
nanotubes have a strong effect on the conductivity of double wall
nanotubes~\cite{souza,shan,dyachkov}. The important parameter is
also a difference of wave vectors $k_{F}$ of the individual
nanotubes.

To compute the influence of a curvature of the surface on the
matrix elements of the secular equation, we used two methods. The
rehybridization of the $\pi$ orbital method was used for the
computation of diagonal matrix elements which define the Fermi
levels of single nanotubes. To compute the nondiagonal matrix
element, we used the curvature tensor $b_{ij}$. In the present
work, we get the same gap as in~\cite{kleiner} which was computed
by the rehybridized method for single wall zig-zag nanotubes. The
gap is by a factor of 4 larger than that computed in the previous
study ~\cite{kane}. The reason is that we get analytically a 4
time bigger term $\delta t_{i}/t$. The curvature of the surface
opens the gap in the zig-zag SWNTs but does not open the gap in
the armchair SWNTs. The Fermi level of the outer shell is about
$0.21$ eV higher than the Fermi level of the inner shell in the
case of $(9,0)-(18,0)$ zig-zag DWNTs. In the case of zig-zag
DWNTs, the curvature does not shift the minimum of the conductance
band and maximum of the valence band of the individual nanotubes.
The result is that these DWNTs are the semiconductor. The
electronic structure of the $(9,0)-(18,0)$ DWNTs in the absence of
the intertube interactions is shown in Fig. 4. Due to the
difference in Fermi levels of individual nanotubes the valence
states are not symmetric to conduction states about the Fermi
level. We have a gap $E_{g}=25$ meV between the valence band of
the outer shell and the conductive band of the inner shell. The
difference in the Fermi levels of individual nanotubes has not
been taken into account in~\cite{lin1}. They have symmetric
valence states to the conduction states, and the energy gap
$E_{g}$ is associated with outer $(18,0)$ nanotubes in the absence
of the intertube interaction. We get a minimum gap between the
valence and conductive band at the points $\sqrt{3}ka/2\simeq \pm
0.05$ and this energy gap has value $E_{g}=90$ meV when the
intertube interactions are imposed. We also compute the energy
gaps of $(8,0)-(16,0)$ and $(10,0)-(20,0)$ zig-zag DWNTs. For
$(8,0)-(16,0)$ DWNTs we get a significantly greater gap than is
predicted by DFT calculations. It is mainly due to difference in
the energy gaps of $(8,0)$ SWNTs. DFT calculations predict energy
gaps $0.6$eV. Quasiparticle corrections open the gap to $1.75$ eV
~\cite{spataru}. A similar gap is predicted in the present paper.
The values of the minimum energy gaps for different types of
zig-zag SWNTs and DWNTs are collected in Tables I and II.

The Fermi level of the outer shell is about $0.23$ eV higher than
the Fermi level of the inner shell for $(5,5)-(10,10)$ armchair
DWNTs. The result is that in the armchair DWNTs part of electrons
from the valence band of the outer shell comes to the conductance
band of the inner shell. The inner shell will have a negative
charge and will have electron conductivity, and the outer shell
will have a positive charge and will have hole conductivity. In
the case of armchair SWNTs, the Fermi points are shifted and the
shift depends on the curvature. Since the $\frac{\alpha}{2\beta}$
is bigger than $\frac{\widetilde{\alpha}}{2\widetilde{\beta}}$,
the Fermi point $k_{F}$ of the outer nanotube is smaller than the
wave vector $k_{F}$ of the Fermi point of the inner nanotube. The
highest occupied state is located above the lowest unoccupied
state in the case of the armchair DWNT. The differences are $0.16$
eV in the symmetric geometry, $0.1$ eV when inner nanotube is
shifted in the direction of the axes and $96$ meV in the case of
the rotational displacement of the inner with respect to the outer
tube. So armchair DWNTs have a semimetallic character. We get the
same character of the conductivity in all computed geometries for
the armchair $(5,5)-(10,10)$ nanotube. It means that the
conductivity does not strongly depend on the relative position of
individual shells. We get similar results, as
in~\cite{saito,kwon}, for the asymmetric geometry of
$(5,5)-(10,10)$ armchair nanotube, but we have the inverse
asymmetry of the electronic spectra. In our model we get the
asymmetry because the Fermi level of the outer nanotube is higher
than the Fermi level of the inner nanotube, and the wave vector
$k_{F}$ of the Fermi level of the outer nanotube is smaller than
the wave vector of the Fermi level of the inner nanotube (Fig. 8).
We get similar results also for $(4,4)-(8,8)$ and $(6,6)-(12,12)$
DWNTs where the highest occupied state is $0.217$ eV above the
lowest unoccupied state in the case of $(4,4)-(8,8)$ DWNTs and
$0.12$ eV in the case of $(6,6)-(12,12)$ DWNTs.

The main reason why there is a difference in the character of the
conductivity of armchair and zig-zag double wall nanotubes is the
absence of the shift of the wave vector $k$ where the individual
zig-zag nanotubes have a minimal gap. So zig-zag DWNTs are
semiconductors. We have a maximum of the valence band of the outer
armchair nanotube higher than a minimum of the conductive band of
the inner nanotube in the armchair DWNTs. There is no energy gap in
armchair nanotubes but there is a shift of $k_{F}$. Those are the
main reasons why the armchair double wall nanotubes with a small
radius are semimetal. We can also conclude that the shift and
rotation of the inner nanotube do not influent largely the main
characteristics of the DWNTs armchair nanotubes; therefore, they are
stable with their semimetallic character. It would be interesting to
test this prediction in experiment.

Generally, we can say that the conductivity depends on the
relative position of the wave vectors  $k$ where the individual
nanotubes have a minimum gap. If there is no shift, the DWNTs are
semiconductors. Zig-zag SWNTs have a minimum gap at point $\Gamma
$. It means that from our prediction all zig-zag DWNTs ought to be
semiconductors. It is partially supported in~\cite{dyachkov}. On
the other hand, if there is a shift in the wave vectors where the
individual nanotubes have a minimal gap depending on the mutual
positions of the Fermi levels and the energy gap width of
individual nanotubes, the DWNT can be semimetal or semiconductor.
The examples are $(5,5)-(10,10)$ DWNTs where the shift is caused
by curvature and $(4,2)-(10,5)$ DWNTs where individual nanotubes
have gap minima at the points $X$ and $\Gamma$~\cite{song}. We
have shown that the difference in the Fermi level energies and
mixing of orbitals localized on the outer and inner nanotubes
cause the charge transfer from outer to inner tubules. We do not
take into account that charge transfer between the outer and inner
tubules create an electric field between these tubules. So not all
electrons can transfer from the outer shell to the inner shell, as
is predicted by the present study. Assumption of this effect can
make a reconstruction of the electronic spectra of DWNTs. This is
important mainly in the DWNTs where the inner nanotube has a very
small diameter. The result can be metallic character of zig-zag
DWNTs with $(7,0)$ and $(5,0)$ inner nanotube, as is predicted
in~\cite{song,Zolyomi,Kurti}. Our calculations predict
semiconducting character of $(9,0)-(18,0)$ DWNTs similarly
to~\cite{kwon,lin1} and contrary to~\cite{Zolyomi}. It ought to be
resolved.

If inner shell has a radius about $7\dot{A}$ and more, the
difference between Fermi energy of the outer and the inner tubules
is small. So the charge transfer is not significant. The lower the
minimum of the $\pi^{*}$ state of the inner nanotube in comparison
with the maximum of the $\pi$ state of the outer nanotube the
bigger charge transfer is. If individual nanotubes have metallic
character, the charge transfer will be greater then in the case of
DWNTs where one or both of the nanotubes are semiconductor. It
means for instance that the charge transfer is smaller in the case
of zig-zag DWNTs than in the case of armchair DWNTs with similar
radius. Charge transfer is from the outer to the inner nanotube
because Fermi level of the outer nanotube is higher then Fermi
level of the inner nanotube.

From Eq.71 we get the following formula for the Fermi wave vector
$k_{F}$ of the armchair SWNT;
\begin{equation}
k_{F}=\frac{2}{a}\arccos\frac{1-\frac{1}{2}\left(\frac{d}{R}\right)^{2}}{2\left(1-\frac{1}{8}\left(\frac{d}{R}\right)^{2}\right)}.
\end{equation}
For a large radius the Fermi wave vector is located at
$k_{F}(R\rightarrow\infty)=2\pi/3a$. As a diameter decreases, the
position of $k_{F}$ shifts from $k_{F}(R\rightarrow\infty)$
towards the bigger wave vectors. The DFT calculations predict the
opposite shift~\cite{Zolyomi1}. Parameter $\alpha $ is smaller
than parameter $\beta$. It means that because of curvature the
hopping integral in the $\vec{\tau}_{1}$ direction is smaller than
the hopping integrals in the $\vec{\tau}_{2}$ and $\vec{\tau}_{3}$
directions. This is the reason why we get the shift of $k_{F}$
towards the bigger wave vector with decreasing of the radius of
the nanotube. We expect that less symmetric DWNTs have no such
stable characteristic when we change a relative position of the
outer and inner nanotubes. The oscillation character of a energy
gap will not exist in the case of less symmetric DWNTs. The
understanding how the rotation of the inner nanotube in different
types of DWNTs influences electronic properties of this type of
nanostructures is needed to design a new type of
nanomotors~\cite{Bailey}.

\vskip 0.2cm ACKNOWLEDGEMENTS --- The authors thank Prof. V.A.
Osipov for helpful discussions and advice. The work was supported by
VEGA grant 2/7056/27 of the Slovak Academy of Sciences and by the
Science and Technology Assistance Agency under contract No. APVV
0509-07.

\newpage

\section{Appendix A}
In a tight-binding approximation for the case of zig-zag tubules we
get the following systems of equations: for the outer shell
\begin{equation}
\epsilon
C_{A_{1}}+H_{A_{1}B_{2}}C_{B_{2}}+H_{A_{1}B_{2}^{`}}C_{B_{2}^{`}}
+H_{A_{1}B_{1}}C_{B_{1}}+\sum _{\lambda}W_{A_{1},\lambda}
C_{\lambda}=EC_{A_{1}},
\end{equation}
where $H_{A_{1}B_{2}}=\gamma_{0}\beta
e^{i\overrightarrow{k}\overrightarrow{\tau_{2}}}$;
$H_{A_{1}B_{2}^{`}}=\gamma_{0}\beta
e^{i\overrightarrow{k}\overrightarrow{\tau_{3}}}$;
$H_{A_{1}B_{1}}=\gamma_{0}
e^{i\overrightarrow{k}\overrightarrow{\tau_{1}}}$.
\begin{equation}
\epsilon
C_{B_{1}}+H_{B_{1}A_{1}}C_{A_{1}}+H_{B_{1}A_{2}^{`}}C_{A_{2}^{`}}+
H_{B_{1}A_{2}}C_{A_{2}}+\sum
_{\lambda}W_{B_{1},\lambda}C_{\lambda}=EC_{B_{1}},
\end{equation}
where $H_{B_{1}A_{1}}=\gamma_{0}
e^{-i\overrightarrow{k}\overrightarrow{\tau_{1}}}$;
$H_{B_{1}A_{2}^{`}}=\gamma_{0}\beta
e^{-i\overrightarrow{k}\overrightarrow{\tau_{2}}}$;
$H_{B_{1}A_{2}}=\gamma_{0}\beta
e^{-i\overrightarrow{k}\overrightarrow{\tau_{3}}}$.
\begin{equation}
\epsilon C_{A_{2}}+H_{A_{2}B_{1}}C_{B_{1}}+H_{A_{2}B_{2}}C_{B_{2}}+
H_{A_{2}B_{1}^{`}}C_{B_{1}^{`}}+\sum
_{\lambda}W_{A_{2},\lambda}C_{\lambda}=EC_{A_{2}},
\end{equation}
where $H_{A_{2}B_{2}}=\gamma_{0}
e^{i\overrightarrow{k}\overrightarrow{\tau_{1}}}$;
$H_{A_{2}B_{1}}=\gamma_{0}\beta
e^{i\overrightarrow{k}\overrightarrow{\tau_{3}}}$;
$H_{A_{2}B_{1}^{`}}=\gamma_{0}\beta
e^{i\overrightarrow{k}\overrightarrow{\tau_{2}}}$.
\begin{equation}
\epsilon
C_{B_{2}}+H_{B_{2}A_{1}}C_{A_{1}}+H_{B_{2}A_{1}^{`}}C_{A_{1}^{`}}+
H_{B_{2}A_{2}}C_{A_{2}}+\sum
_{\lambda}W_{B_{2},\lambda}C_{\lambda}=EC_{B_{2}},
\end{equation}
where $H_{B_{2}A_{1}}=\gamma_{0}\beta
e^{-i\overrightarrow{k}\overrightarrow{\tau_{2}}}$;
$H_{B_{2}A_{1}^{`}}=\gamma_{0}\beta
e^{-i\overrightarrow{k}\overrightarrow{\tau_{3}}}$;
$H_{B_{2}A_{2}}=\gamma_{0}
e^{-i\overrightarrow{k}\overrightarrow{\tau_{1}}}$.
\begin{equation}
\epsilon
C_{B_{1}^{`}}+H_{B_{1}^{`}A_{2}}C_{A_{2}}+H_{B_{1}^{`}A_{2}^{`}}C_{A_{2}^{`}}
+H_{B_{1}^{`}A_{1}^{`}}C_{A_{1}^{`}}+\sum
_{\lambda}W_{B_{1}^{'},\lambda}C_{\lambda}=EC_{B_{1}^{`}},
\end{equation}
where $H_{B_{1}^{`}A_{2}}=\gamma_{0}\beta
e^{-i\overrightarrow{k}\overrightarrow{\tau_{2}}}$;
$H_{B_{1}^{`}A_{2}^{`}}=\gamma_{0}\beta
e^{-i\overrightarrow{k}\overrightarrow{\tau_{3}}}$;
$H_{B_{1}^{`}A_{1}^{`}}=\gamma_{0}
e^{-i\overrightarrow{k}\overrightarrow{\tau_{1}}}$.
\begin{equation}
\epsilon
C_{A_{2}^{`}}+H_{A_{2}^{`}B_{1}^{`}}C_{B_{1}^{`}}+H_{A_{2}^{`}B_{2}^{`}}C_{B_{2}^{`}}
+H_{A_{2}^{`}B_{1}}C_{B_{1}}+\sum
_{\lambda}W_{A_{2}^{'},\lambda}C_{\lambda}=EC_{A_{2}^{`}},
\end{equation}
where $H_{A_{2}^{`}B_{1}^{`}}=\gamma_{0}\beta
e^{i\overrightarrow{k}\overrightarrow{\tau_{3}}}$;
$H_{A_{2}^{`}B_{2}^{`}}=\gamma_{0}
e^{i\overrightarrow{k}\overrightarrow{\tau_{1}}}$;
$H_{A_{2}^{`}B_{1}}=\gamma_{0}\beta
e^{i\overrightarrow{k}\overrightarrow{\tau_{2}}}$.
\begin{equation}
\epsilon
C_{B_{2}^{`}}+H_{B_{2}^{`}A_{2}^{`}}C_{A_{2}^{`}}+H_{B_{2}^{`}A_{1}^{`}}C_{A_{1}^{`}}
+H_{B_{2}^{`}A_{1}}C_{A_{1}}+\sum
_{\lambda}W_{B_{2}^{`},\lambda}C_{\lambda}=EC_{B_{2}^{`}},
\end{equation}
where $H_{B_{2}^{`}A_{2}^{`}}=\gamma_{0}
e^{-i\overrightarrow{k}\overrightarrow{\tau_{1}}}$;
$H_{B_{2}^{`}A_{1}^{`}}=\gamma_{0}\beta
e^{-i\overrightarrow{k}\overrightarrow{\tau_{2}}}$;
$H_{B_{2}^{`}A_{1}}=\gamma_{0}\beta
e^{-i\overrightarrow{k}\overrightarrow{\tau_{3}}}$.
\begin{equation}
\epsilon
C_{A_{1}^{`}}+H_{A_{1}^{`}B_{1}^{`}}C_{B_{1}^{`}}+H_{A_{1}^{`}B_{2}}C_{B_{2}}
+H_{A_{1}^{`}B_{2}^{`}}C_{B_{2}^{`}}++\sum
_{\lambda}W_{A_{1}^{`},\lambda}C_{\lambda}=EC_{A_{1}^{`}},
\end{equation}
where $H_{A_{1}^{`}B_{1}^{`}}=\gamma_{0}
e^{i\overrightarrow{k}\overrightarrow{\tau_{1}}}$;
$H_{A_{1}^{`}B_{2}}=\gamma_{0}\beta
e^{i\overrightarrow{k}\overrightarrow{\tau_{3}}}$;
$H_{A_{1}^{`}B_{2}^{`}}=\gamma_{0}\beta
e^{i\overrightarrow{k}\overrightarrow{\tau_{2}}}$. Here $\lambda$
denotes the atoms of the unitary cell localized on the inner shell.
Now we write down the equations for the inner shell in the case of
zigzag nanotubes.
\begin{equation}
\tilde{\epsilon} C_{A}+H_{AB}C_{B}+H_{AB^{`}}C_{B^{`}}+\sum
_{\lambda}W_{A,\lambda}C_{\lambda}=EC_{A},
\end{equation}
where $H_{AB}=\gamma_{0}
e^{i\overrightarrow{k}\overrightarrow{\tau_{1}}}$;
$H_{AB^{`}}=\gamma_{0}\widetilde{\beta}(
e^{i\overrightarrow{k}\overrightarrow{\tau_{2}}}+
e^{i\overrightarrow{k}\overrightarrow{\tau_{3}}})$.
\begin{equation}
\tilde{\epsilon} C_{B}+H_{BA}C_{A}+H_{BA^{`}}C_{A^{`}}+\sum
_{\lambda}W_{B,\lambda}C_{\lambda}=EC_{B},
\end{equation}
where $H_{BA}=\gamma_{0}
e^{-i\overrightarrow{k}\overrightarrow{\tau_{1}}}$;
$H_{BA^{`}}=\gamma_{0}\widetilde{\beta}(
e^{-i\overrightarrow{k}\overrightarrow{\tau_{2}}}+
e^{-i\overrightarrow{k}\overrightarrow{\tau_{3}}})$.
\begin{equation}
\tilde{\epsilon}
C_{A^{`}}+H_{A^{`}B}C_{B}+H_{A^{`}B^{`}}C_{B^{`}}+\sum
_{\lambda}W_{A^{`},\lambda}C_{\lambda}=EC_{A^{`}},
\end{equation}
where $H_{A^{`}B^{`}}=\gamma_{0}
e^{i\overrightarrow{k}\overrightarrow{\tau_{1}}}$;
$H_{A^{`}B}=\gamma_{0}\widetilde{\beta}(
e^{i\overrightarrow{k}\overrightarrow{\tau_{2}}}+
e^{i\overrightarrow{k}\overrightarrow{\tau_{3}}})$.
\begin{equation}
\tilde{\epsilon}
C_{B^{`}}+H_{B^{`}A}C_{A}+H_{B^{`}A^{`}}C_{A^{`}}+\sum
_{\lambda}W_{B^{`},\lambda}C_{\lambda}=EC_{B^{`}},
\end{equation}
where $H_{B^{`}A^{`}}=\gamma_{0}
e^{-i\overrightarrow{k}\overrightarrow{\tau_{1}}}$;
$H_{B^{`}A}=\gamma_{0}\widetilde{\beta}(
e^{-i\overrightarrow{k}\overrightarrow{\tau_{2}}}+
e^{-i\overrightarrow{k}\overrightarrow{\tau_{3}}})$ and $\lambda$
denotes the atoms of the unitary cell localized on the outer shell.

\section{Appendix B}
In a tight-binding approximation for the case of armchair tubules we
get the following systems of equations:
for the outer shell
\begin{equation}
\epsilon
C_{A_1}+H_{A_{1}B_{1}}C_{B_{1}}+H_{A_{1}B_{2}^{`}}C_{B_{2}^{`}}+
\sum_{\lambda} W_{A_{1},\lambda}C_{\lambda}=EC_{A_{1}},
\end{equation}
where $H_{A_{1}B_{1}}=\gamma_{0}\alpha
e^{i\overrightarrow{k}\overrightarrow{\tau_{1}}}$;
$H_{A_{1}B_{2}^{`}}=\gamma_{0}\beta(
e^{i\overrightarrow{k}\overrightarrow{\tau_{2}}}+
e^{i\overrightarrow{k}\overrightarrow{\tau_{3}}})$.
\begin{equation}
\epsilon
C_{B_1}+H_{B_{1}A_{1}}C_{A_{1}}+H_{B_{1}A_{2}}C_{A_{2}}+\sum_{\lambda}
W_{B_{1},\lambda}C_{\lambda}=EC_{B_{1}},
\end{equation}
where $H_{B_{1}A_{1}}=\gamma_{0}\alpha
e^{-i\overrightarrow{k}\overrightarrow{\tau_{1}}}$;
$H_{B_{1}A_{2}}=\gamma_{0}\beta(
e^{-i\overrightarrow{k}\overrightarrow{\tau_{2}}}+
e^{-i\overrightarrow{k}\overrightarrow{\tau_{3}}})$.
\begin{equation}
\epsilon
C_{A_2}+H_{A_{2}B_{2}}C_{B_{2}}+H_{A_{2}B_{1}}C_{B_{1}}+\sum_{\lambda}
W_{A_{2},\lambda}C_{\lambda}=EC_{A_{2}},
\end{equation}
where $H_{A_{2}B_{2}}=\gamma_{0}\alpha
e^{i\overrightarrow{k}\overrightarrow{\tau_{1}}}$;
$H_{A_{2}B_{1}}=\gamma_{0}\beta(
e^{i\overrightarrow{k}\overrightarrow{\tau_{2}}}+
e^{i\overrightarrow{k}\overrightarrow{\tau_{3}}})$.
\begin{equation}
\epsilon
C_{B_2}+H_{B_{2}A_{1}^{`}}C_{A_{1}^{`}}+H_{B_{2}A_{2}}C_{A_{2}}+\sum_{\lambda}
W_{B_{2},\lambda}C_{\lambda}=EC_{B_{2}},
\end{equation}
where $H_{B_{2}A_{2}}=\gamma_{0}\alpha
e^{-i\overrightarrow{k}\overrightarrow{\tau_{1}}}$;
$H_{B_{2}A_{1}^{`}}=\gamma_{0}\beta(
e^{-i\overrightarrow{k}\overrightarrow{\tau_{2}}}+
e^{-i\overrightarrow{k}\overrightarrow{\tau_{3}}})$.
\begin{equation}
\epsilon
C_{A_1^{`}}+H_{A_{1}^{`}B_{2}}C_{B_{2}}+H_{A_{1}^{`}B_{1}^{`}}C_{B_{1}^{`}}+\sum_{\lambda}
W_{A_{1}^{`},\lambda}C_{\lambda}=EC_{A_{1}^{`}},
\end{equation}
where $H_{A_{1}^{`}B_{1}^{`}}=\gamma_{0}\alpha
e^{i\overrightarrow{k}\overrightarrow{\tau_{1}}}$;
$H_{A_{1}^{`}B_{2}}=\gamma_{0}\beta(
e^{i\overrightarrow{k}\overrightarrow{\tau_{2}}}+
e^{i\overrightarrow{k}\overrightarrow{\tau_{3}}})$.
\begin{equation}
\epsilon
C_{B_1^{`}}+H_{B_{1}^{`}A_{1}^{`}}C_{A_{1}^{`}}+H_{B_{1}^{`}A_{2}^{`}}C_{A_{2}^{`}}+\sum_{\lambda}
W_{B_{1}^{`},\lambda}C_{\lambda}=EC_{B_{1}^{`}},
\end{equation}
where $H_{B_{1}^{`}A_{1}^{`}}=\gamma_{0}\alpha
e^{-i\overrightarrow{k}\overrightarrow{\tau_{1}}}$;
$H_{B_{1}^{`}A_{2}^{`}}=\gamma_{0}\beta(
e^{-i\overrightarrow{k}\overrightarrow{\tau_{2}}}+
e^{-i\overrightarrow{k}\overrightarrow{\tau_{3}}})$.
\begin{equation}
\epsilon
C_{A_2^{`}}+H_{A_{2}^{`}B_{1}^{`}}C_{B_{1}^{`}}+H_{A_{2}^{`}B_{2}^{`}}C_{B_{2}^{`}}+\sum_{\lambda}
W_{B_{2}^{`},\lambda}C_{\lambda}=EC_{B_{2}^{`}},
\end{equation}
where $H_{A_{2}^{`}B_{2}^{`}}=\gamma_{0}\alpha
e^{i\overrightarrow{k}\overrightarrow{\tau_{1}}}$;
$H_{A_{2}^{`}B_{1}^{`}}=\gamma_{0}\beta(
e^{i\overrightarrow{k}\overrightarrow{\tau_{2}}}+
e^{i\overrightarrow{k}\overrightarrow{\tau_{3}}})$.
\begin{equation}
\epsilon
C_{B_2^{`}}+H_{B_{2}^{`}A_{1}}C_{A_{1}}+H_{B_{2}^{`}A_{2}^{`}}C_{A_{2}^{`}}+\sum_{\lambda}
W_{B_{2}^{`},\lambda}C_{\lambda}=EC_{B_{2}^{`}},
\end{equation}
where $H_{B_{2}^{`}A_{2}^{`}}=\gamma_{0}\alpha
e^{-i\overrightarrow{k}\overrightarrow{\tau_{1}}}$;
$H_{B_{2}^{`}A_{1}}=\gamma_{0}\beta(
e^{-i\overrightarrow{k}\overrightarrow{\tau_{2}}}+
e^{-i\overrightarrow{k}\overrightarrow{\tau_{3}}})$. Here $\lambda$
denotes the atoms of the unitary cell localized on the inner shell.
The equations for the inner shell can be expressed in the form:
\begin{equation}
\tilde{\epsilon}
C_{A}+H_{AB^{`}}C_{B^{`}}+H_{AB}C_{B}+\sum_{\lambda}W_{A,\lambda}C_{\lambda}
=EC_{A},
\end{equation}
where $H_{AB}=\gamma_{0}\widetilde{\alpha}
e^{i\overrightarrow{k}\overrightarrow{\tau_{1}}}$;
$H_{AB^{`}}=\gamma_{0}\widetilde{\beta}(
e^{i\overrightarrow{k}\overrightarrow{\tau_{2}}}+
e^{i\overrightarrow{k}\overrightarrow{\tau_{3}}})$.
\begin{equation}
\tilde{\epsilon}
C_{B}+H_{BA}C_{A}+H_{BA^{`}}C_{A^{`}}+\sum_{\lambda}W_{B,\lambda}C_{\lambda}=EC_{B},
\end{equation}
where $H_{BA}=\gamma_{0}\widetilde{\alpha}
e^{-i\overrightarrow{k}\overrightarrow{\tau_{1}}}$;
$H_{BA^{`}}=\gamma_{0} \widetilde{\beta}(
e^{-i\overrightarrow{k}\overrightarrow{\tau_{2}}}+
e^{-i\overrightarrow{k}\overrightarrow{\tau_{3}}})$.
\begin{equation}
\tilde{\epsilon}
C_{A^{`}}+H_{A^{`}B}C_{B}+H_{A^{`}B^{`}}C_{B^{`}}+\sum_{\lambda}W_{A^{`},\lambda}C_{\lambda}=EC_{A^{`}},
\end{equation}
where $H_{A^{`}B^{`}}=\gamma_{0}\widetilde{\alpha}
e^{i\overrightarrow{k}\overrightarrow{\tau_{1}}}$;
$H_{A^{`}B}=\gamma_{0}\widetilde{\beta}(
e^{i\overrightarrow{k}\overrightarrow{\tau_{2}}}+
e^{i\overrightarrow{k}\overrightarrow{\tau_{3}}})$.
\begin{equation}
\tilde{\epsilon}
C_{B^{`}}+H_{B^{`}A}C_{A}+H_{B^{`}A^{`}}C_{A^{`}}+\sum_{\lambda}W_{B^{`},\lambda}C_{\lambda}=EC_{B^{`}},
\end{equation}
where $H_{BA^{`}}=\gamma_{0}\widetilde{\alpha}
e^{-i\overrightarrow{k}\overrightarrow{\tau_{1}}}$;
$H_{B^{`}A}=\gamma_{0}\widetilde{\beta}(
e^{-i\overrightarrow{k}\overrightarrow{\tau_{2}}}+
e^{-i\overrightarrow{k}\overrightarrow{\tau_{3}}})$. Here $\lambda$
denotes the atoms of the unitary cell localized on the outer shell.

\end{document}